\documentclass[graybox]{svmult}

\usepackage{mathptmx}       % selects Times Roman as basic font
\usepackage{helvet}         % selects Helvetica as sans-serif font
\usepackage{courier}        % selects Courier as typewriter font
\usepackage{type1cm}        % activate if the above 3 fonts are
\usepackage{makeidx}         % allows index generation
\usepackage{graphicx}        % standard LaTeX graphics tool
\usepackage{multicol}        % used for the two-column index
\usepackage[bottom]{footmisc}% places footnotes at page bottom

\makeindex             % used for the subject index
\def\p{\partial}
\def\ep{\varepsilon}

\def\sm{\smallskip}
\def\no{\nonumber}

\newcommand{\bea}{\begin{eqnarray}}
\newcommand{\eea}{\end{eqnarray}}
\newcommand{\<}{\langle}
\renewcommand{\>}{\rangle}

\begin{document}

\title*{Quantum Criticality  via Magnetic Branes}
\author{Eric D'Hoker and Per Kraus}
\institute{Eric D'Hoker \at Department of Physics and Astronomy, University of California,
Los Angeles, CA 90095, USA, \email{dhoker@physics.ucla.edu}
\and Per Kraus \at Department of Physics and Astronomy, University of California,
Los Angeles, CA 90095, USA,  \email{pkraus@physics.ucla.edu}}
\maketitle

\abstract{Holographic methods are used to investigate the low temperature limit, including
quantum critical behavior, of strongly coupled 4-dimensional gauge theories in the presence
of an external magnetic field, and finite charge density. In addition to the metric, the dual gravity theory
contains a Maxwell field with Chern-Simons coupling. In the absence of charge,
the magnetic field induces an RG flow to an infrared AdS$_3 \times {\bf R}^2$ geometry, which is
dual to a 2-dimensional CFT representing strongly interacting fermions in the lowest Landau level.
Two asymptotic Virasoro algebras and one chiral Kac-Moody algebra arise as {\sl emergent symmetries}
in the IR.  Including a nonzero charge density reveals a quantum critical point when the magnetic field reaches
a critical value whose scale is set by the charge density. The critical theory is probed by the study of
long-distance correlation functions of the boundary stress tensor and current.  All  quantities of major physical interest in this system, such as critical exponents and scaling functions, can be computed analytically.
We also study an asymptotically AdS$_6$ system whose magnetic field induced quantum critical point is governed by a IR Lifshitz geometry,  holographically dual to a  D=2+1  field theory.  The behavior of these holographic theories shares important similarities with that of real world quantum critical systems obtained by tuning a magnetic field,
and may be relevant to materials such as Strontium Ruthenates.}

%%%%%%%%%%%%%%%%%%%%%%%%%%%%%%%%%%%%%%%%%%%%%%%%%%%%
%%%%%%%%%%%%%%%%%%%%%%%%%%%%%%%%%%%%%%%%%%%%%%%%%%%%
\section{Introduction}
\label{sec:1}
%%%%%%%%%%%%%%%%%%%%%%%%%%%%%%%%%%%%%%%%%%%%%%%%%%%%
%%%%%%%%%%%%%%%%%%%%%%%%%%%%%%%%%%%%%%%%%%%%%%%%%%%%

A statistical mechanics system undergoes a quantum phase transition when its ground state
suffers a macroscopic rearrangement as an external parameter is varied. While a quantum
phase transition takes place strictly at zero temperature, its presence governs quantum critical
behavior in a small region of low temperature surrounding the quantum critical point. The existence
of such a quantum critical region is believed to influence physics also at intermediate temperatures,
and to have relevance to the phase structure of high-$T_c$ superconductors and strange metals
\cite{Sachdev}.

\sm

Holography provides concrete tools for studying non-Abelian gauge dynamics
in terms of classical solutions to Einstein's equations of gravity, provided the number of
colors $N$ and the `t Hooft coupling $\lambda = N g_{\rm \scriptstyle YM}^2$ are both large
\cite{Maldacena:1997re,Gubser:1998bc,Witten:1998qj}.
The classic example of this gauge/gravity duality relates ${\cal N}=4$ super Yang-Mills theory in
3+1  dimensional Minkowski space-time to Type IIB supergravity on $AdS_5 \times S^5$.
Global symmetries match under the duality: the space-time isometry group $SO(4,2) \times SO(6)$
on the gravity side maps to the conformal group $SO(4,2)$ and the R-symmetry group $SU(4)\sim SO(6)$
on the gauge theory side, while the number of supersymmetries is the maximal allowed 32 on both
sides  (for general reviews on the AdS/CFT correspondence, see for example
\cite{Aharony:1999ti, Klebanov:2000me, D'Hoker:2002aw, Polchinski:2010hw}).

\sm

The gauge/gravity duality continues to apply to systems with fewer or no supersymmetries,
and with broken conformal and Poincar\'e invariance. For example,
renormalization group flow away from a conformal invariant gauge theory is dual to a space-time
which is {\sl asymptotically} $AdS_5$ but deviates from $AdS_5$ away from the asymptotic limit.
The dual to a gauge theory at finite temperature $T$ is a space-time containing a black hole or
black brane whose Hawking temperature is $T$. A charge density or chemical potential,
and a background magnetic field  may all be incorporated in the gauge theory as well
via  precise dual gravity prescriptions, thereby setting the stage for applying holographic
methods to a wide variety of interesting strongly coupled systems in statistical mechanics.
Reviews on the applications of holographic methods to condensed matter problems may be found in
\cite{Hartnoll:2009sz,Herzog:2009xv}.

\sm

Whether holography will ever be able to model reliably the condensed state of a specific
compound remains to be seen. What has become clear, however, is that gauge/gravity duality
can provide quantitative information on universal behavior, such as critical phenomena, critical
exponents, transport properties, and the like for certain classes of strongly coupled systems.
One of the first examples derived in this spirit is the ratio of the shear viscosity to the entropy
density \cite{Policastro:2001yc}; a more recent one gives a bound on the specific heat
exponent \cite{Ogawa:2011bz}.

\sm

In the present paper, we shall review recent holographic investigations into the critical
behavior of gauge theory at finite temperature $T$, with electric charge density $\rho$,
and subjected to an external magnetic field $B$. Supersymmetry will play no significant role here,
as the gauge theory in question may be supersymmetric or not. We shall be interested in
the thermodynamic properties of this system especially at low temperatures,
as well as in the behavior of correlators  at long distances.

\begin{figure}[htb]
\sidecaption[t]
\includegraphics[scale=.5]{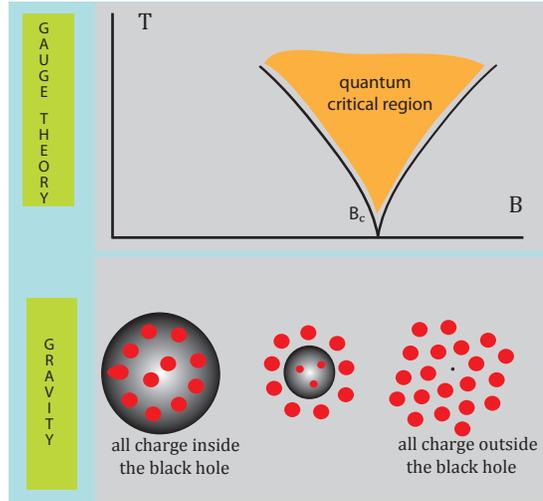}
\caption{The holographic picture dual to a meta-magnetic quantum phase transition is given by
the gradual expulsion of electric charge from the region interior to the event horizon of a black brane
to the region outside the horizon. The quantum critical point corresponds to the transition point
at which  all electric charge resides outside the horizon of the black brane.}
\label{fig:1}
\end{figure}

The holographic dual to this system, in the large $N$ and large $\lambda$ approximations,
is provided by a theory of gravity in 4+1 space-time dimensions, plus an Abelian gauge field.
Anomalies in the gauge current of the 3+1-dimensional field theory side force the presence of
a Chern-Simons term in the 4+1-dimensional theory. The CS coupling is the only dimensionless
free parameter in the theory, and the holographic dynamics will depend crucially on its value.
All our results will be derived using this holographic model \cite{D'Hoker:2009mm,D'Hoker:2009bc}.

\sm

Using holographic methods, a rich structure is found which exhibits quantum critical
behavior \cite{D'Hoker:2010rz,D'Hoker:2010ij,D'Hoker:2012ej},
and the emergence of a 1+1-dimensional CFT in IR correlators \cite{D'Hoker:2010hr,D'Hoker:2011xw}.
Specifically, the system  exhibits a quantum phase transition as the magnetic field $B$ crosses
a critical value $B_c$, the scale for which is set by the charge density $\rho$ by $B_c \sim \rho ^{2/3}$.
Quantum critical behavior governs a region, depicted schematically in figure \ref{fig:1},
where temperature is the largest scale. The mechanism underlying this transition on the gravity
side is also illustrated in figure \ref{fig:1}: it is driven by the expulsion of electric charge from
within the horizon to the outside. More specifically, for vanishing magnetic field, the gravity
solution is the AdS Reissner-Nordstrom black brane which has non-zero charge density and
entropy density at $T=0$. As $B$ is increased electric charge gets expelled from within the
black brane horizon to the outside, up till $B=B_c$ at which value the black brane carries
no more charge or entropy density.  This charge expulsion mechanism is realized in other
holographic systems as well \cite{Hartnoll:2011pp}.
Other  examples of magnetic field driven  holographic phase transitions  include
\cite{Lifschytz:2009sz,Jensen:2010vd}.

%%%%%%%%%%%%%%%%%%%%%%%%%%%%%%%%%%%%%%%%%%%%%%%%%%%%
%%%%%%%%%%%%%%%%%%%%%%%%%%%%%%%%%%%%%%%%%%%%%%%%%%%%
\section{Basic Gauge Theory Dynamics}
\label{sec:2}
%%%%%%%%%%%%%%%%%%%%%%%%%%%%%%%%%%%%%%%%%%%%%%%%%%%%
%%%%%%%%%%%%%%%%%%%%%%%%%%%%%%%%%%%%%%%%%%%%%%%%%%%%

Before embarking on the study of strongly coupled gauge theory with the tools of holography,
we shall summarize here some basic results on the dynamics of gauge theory in the presence
of an external magnetic field.

\subsection{Effective low energy degrees of freedom}
\label{sec:2a}

In the presence of a constant magnetic field $B$, the energy levels of
massless free bosons and fermions of electric charge $q$ are given as follows,
\bea
\hbox{bosons} ~~ & \hskip 0.8in & E = \sqrt{p^2 + (2n+1) |qB|}  \hskip 0.4in n=0,1,2,\cdots
\no \\
\hbox{fermions} & \hskip 0.8in & E = \sqrt{p^2 + 2n |qB|}  \hskip 0.7in n=0,1,2,\cdots
\no
\eea
Here $p$ is the momentum component parallel to the magnetic field $B$. The energy levels
for bosons and fermions clearly do not match, so that supersymmetry is manifestly broken.
Simple modifications in which supersymmetry is restored do exist, however, and were studied
in \cite{Almuhairi:2011ws}.

\sm

For large $B$, only fermions in the lowest Landau level remain massless.
More precisely, the fermions in the lowest Landau level are 1+1-dimensional Weyl fermions,
moving along the direction of the magnetic field, with momentum $p$,
their chirality being correlated with their charge,
\bea
Bq >0 & \hskip 0.6in p >0 & \hskip 0.6in \hbox{field of right-movers} ~~ \psi _R
\no \\
Bq <0 & \hskip 0.6in p <0 & \hskip 0.6in \hbox{field of left-movers} ~ ~~ \psi _L
\no
\eea
All higher fermion levels and all boson levels acquire a large effective mass and will decouple
from the spectrum.

\sm

In ${\cal N}=4$ super-Yang-Mills theory, the operator of charges $Q$ to which
the magnetic field couples takes values in the R-symmetry algebra $SU(4)_R$. Although
the gauge fields $A_\mu$, gauginos $\lambda_\alpha $, and scalars $\phi$ are now all
strongly interacting, the above decoupling of charged fields will persist. Thus, the low energy
effective degrees of freedom will be Weyl fermions $\psi _L, \psi _R$, whose coupling
is induced by the remaining neutral gauge dynamics of the $A_\mu$ fields, and any remaining
components of $\lambda$ and $\phi$ which are neutral under $Q$. This system of Weyl fermions
should underlie an interacting conformal field theory in 1+1 dimensions. Any {\sl asymmetry}
in the spectrum of charges $Q$ will give rise to the {\sl chiral magnetic effect}, the observation
of which is being considered in heavy ion collision experiments at RHIC \cite{Kharzeev:2007jp}.

\sm

\begin{figure}[htb]
\sidecaption[t]
\includegraphics[scale=.5]{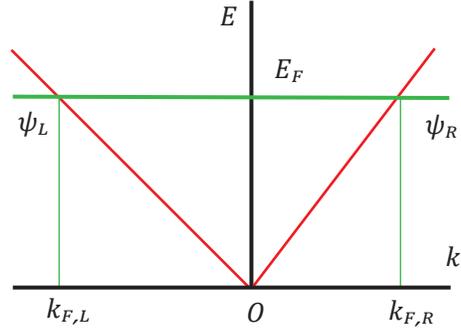}
\caption{At finite charge density, fermion levels fill up to a Fermi energy $E_F$,
corresponding to Fermi momenta $k_{F,L}$ and $k_{F,R}$. The low temperature
degrees of freedom are chiral fermions $\psi _L$ and $\psi _R$ which reside at the
Fermi ``surface", consisting here of only two points $k = k_{F,L}$ and $k = k_{F,R}$.
In the charge symmetric case we have $k_{F,L}=-k_{F,R}$.}
\label{fig:2}
\end{figure}

At finite density and low or zero temperature, fermionic levels will fill up to a certain Fermi energy $E_F$.
For the effective 1+1-dimensional CFT discussed here, the Fermi surface consists of just two points
corresponding to two values of the Fermi momentum. The left- and right-movers
$\psi_L$ and $\psi_R$ live separately at $k=k_{F,L}$ and $k=k_{F,R}$ respectively.
If the spectrum of charges is symmetric then $k_{F,L}=-k_{F,R}$, as is illustrated
in figure \ref{fig:2}.  But in general, $k_{F,L} \neq -k_{F,R}$, and the ground state of the
system will carry a nonzero total momentum.
As the charge density $\rho$ is being increased, $E_F$ will increase.
When $E_F$ reaches the next Landau level, or the energies characteristic of the fully
3-dimensional excitations, a large number of degrees of freedom are being excited, and
we may expect a quantum phase transition, at a critical charge density $\rho_c$,
whose scale is set by the only possible dimensional scale
in the problem, namely $\rho_c \sim B^{3/2}$.

\subsection{Luttinger liquids}
\label{sec:2b}

The standard quantum field theory approach to systems of 1+1-dimensional interacting chiral fermions
is provided by the Luttinger approach to quantum liquids. One begins by identifying the excitations near
the Fermi surface, in this case the Weyl fermions $\psi _L, \psi _R$ introduced above.
The Hamiltonian consists of bilinear terms which result from linearization around the Fermi surface,
as well as all possible local four-fermi interactions compatible with the symmetries of the system,
\bea
\label{2b1}
H_{\rm int} = g_2 (\psi ^\dagger _L \psi _L) (\psi ^\dagger _R \psi _R)
+ {g_4\over 2} (\psi ^\dagger _L \psi _L) ^2 +{ g_4\over 2} (\psi ^\dagger _R \psi _R) ^2
\eea
Although the system was first solved in terms of fermionic fields by Dzyaloshinski and Larkin,
modern methods based on bosonization provide a powerful reformulation in terms of two
non-interacting boson fields \cite{Giamarchi}. A key thermodynamic relation for the entropy
density $s_{\rm gauge}$ as a function of the temperature $T$ is given by,
\bea
\label{2b2}
s_{\rm gauge} = { \pi \over 3v} \, T
\hskip 1in
v = v_F \sqrt{ \left(1+{g_4\over 2\pi v_F} \right)^2 - \left({g_2\over 2\pi v_F}\right)^2}
\eea
where $v_F$ is the Fermi velocity, and $v$ the actual velocity of the chiral excitations.
Correlators may be obtained as well. For example, the two-point function of the charge
density $\rho (x)$ takes the following form,
\bea
\label{2b3}
\< \rho (x) \rho(0)\> = { c_0 \over x^2} + c_\Delta {\cos ( 2 k_F x) \over x^{2 \Delta}} + \cdots
\eea
where $c_0$ and $c_\Delta$ are constants, and $\Delta$ is the scaling dimension of the
lowest dimensional operator which exchanges charge between $\psi _L$ and $\psi _R$.

%%%%%%%%%%%%%%%%%%%%%%%%%%%%%%%%%%%%%%%%%%%%%%%%%%%%
%%%%%%%%%%%%%%%%%%%%%%%%%%%%%%%%%%%%%%%%%%%%%%%%%%%%
\section{Holographic Dual set-up}
\label{sec:3}
%%%%%%%%%%%%%%%%%%%%%%%%%%%%%%%%%%%%%%%%%%%%%%%%%%%%
%%%%%%%%%%%%%%%%%%%%%%%%%%%%%%%%%%%%%%%%%%%%%%%%%%%%

In this section, we shall discuss the basic set up for the holographic dual in the supergravity
approximation. Since we shall concentrate on thermodynamics, as well as on correlators of energy
density, momentum density, and charge density, we may limit the quantum field operators
to the stress tensor ${\cal T}^{\mu \nu}$ and the Maxwell current ${\cal J}^\mu$ of the 3+1-dimensional gauge
theory. The holographic dual fields to these operators are respectively the metric $g_{MN}$ and the
Maxwell field $A_M$ of the 4+1-dimensional Einstein-Maxwell-Chern-Simons theory with
action,\footnote{Einstein indices $\mu, \nu =0,1,2,3$ will be used in 3+1-dimensions, while Einstein
indices $M,N=0,1,2,3,4$ will be used in 4+1-dimensions.
Our conventions are $g=-\det (g_{MN})$, as well as
$R^L{}_{MNK} = \p_K \Gamma^L_{MN} - \p_N \Gamma^L_{MK}
+\Gamma^P_{MN} \Gamma^L_{KP} - \Gamma^P_{MK}\Gamma^L_{NP}$
with $R_{MN} = R^L{}_{MLN}$ and $R=g^{MN} R_{MN}$.  }
\bea
\label{3a1}
S = - { 1 \over 16 \pi G_5} \int d^5 x \, \sqrt{g} \left ( R - {12 \over \ell ^2}  + F_{MN} F^{MN} \right )
+{ k \over 12 \pi G_5} \int A \wedge F \wedge F
\eea
Boundary as well as counterterm contributions to the action have not been exhibited here.
Furthermore, $G_5$ is Newton's constant in 4+1 dimensions, $-12/\ell ^2$ is the cosmological constant,
and $k$ is the dimensionless Chern-Simons coupling.
The anomaly of the chiral current ${\cal J}^\mu$ in the gauge theory is proportional to $k$, and we
have $\p_\mu {\cal J}^\mu \propto k \, {\bf E} \cdot {\bf B}$. The action is invariant under simultaneous reversal
of the sign of $A$ and $k$, allowing us to restrict attention to $k \geq 0$, without loss of generality.

\sm

For the special value $k=k_{\rm \scriptstyle susy} = 2/\sqrt{3}$, the action $S$ coincides with
the bosonic part of minimal
supergravity in 4+1 dimensions, and as such corresponds to a consistent truncation of all
supersymmetric asymptotically $AdS_5$ compactifications of either Type IIB supergravity or
M-theory \cite{Gauntlett:2007ma}. Here, however, we shall leave $k$ a free parameter, and
investigate the phase diagram as a function of $k$.

\subsection{Field equations and structure of the solutions}
\label{sec:3b}

The Bianchi identity is $dF=0$, while the field equations are given as follows,
\bea
\label{3b1}
0 & = &  d * F + k F \wedge F
\no \\
R_{MN} & = & {4 \over \ell^2} g_{MN} + {1 \over 3} g_{MN} F^{PQ} F_{PQ} - 2 F_{MP} F_N{}^P
\eea
For vanishing Maxwell field $F=0$, the field equations admit the $AdS_5$ solution of radius $\ell$.
Henceforth, we shall set $\ell=1$. Denoting the coordinates of 4-dimensional space-time by
$x^\mu = (t, x_1, x_2, x_3)$, and the holographic coordinate by $r$, the $AdS_5$ solution takes
the form,
\bea
\label{3b2}
ds^2 =g_{MN} dx^M dx^N=  {dr^2 \over 4r^2} + 2r \Big ( -dt^2 +  dx_1^2 +  dx_2^2+  dx_3^2 \Big )
\eea
Introducing a constant uniform magnetic field $B$ along the direction $x_3$, and anticipating also the
inclusion of finite temperature $T$, and constant uniform charge density $\rho$, we see that the
symmetries to be imposed on the solutions should include,
\begin{enumerate}
\itemsep= 0in
\item Translation invariance in the coordinates $t, x_1, x_2, x_3$;
\item Rotation invariance in the $x_1,x_2$-plane.
\end{enumerate}
The most general Ansatz consistent with these requirements is given by,
\bea
\label{3b3}
F & = & B dx_1 \wedge dx_2 + E dr \wedge dt - P dr \wedge dx_3 + \tilde P dt \wedge dx_3
\no \\
ds^2 & = & f^{-1} dr^2  + Mdt^2 + 2L dt dx_3 + N dx_3^2 + K (dx_1^2 +  dx_2^2)
\eea
where all coefficient functions $B, E, P, \tilde P, f, K, L, M, N$ depend only on $r$.
In view of the Bianchi identities, $B$ and $\tilde P$ must be independent of $r$, and
in view of the field equations, we have $\tilde P=0$. This constant $B$ is nothing but the
constant magnetic background field. Finally, the residual reparametrization
invariance in the variable $r$ allows us to choose a coordinate $r$ such that
\bea
\label{3b4}
f = L^2 -MN
\eea
a choice which will prove convenient throughout.

\subsection{Boundary stress tensor and current}
\label{sec:3c}

We will be considering asymptotically AdS$_5$ solutions, for which the metric
and gauge field admit a Fefferman-Graham expansion \cite{FG}.
Introducing a radial coordinate $\rho$, defined such that the AdS$_5$ boundary is
located at $\rho=\infty$, the Fefferman-Graham gauge choice puts the fields in the following form,
\bea
\label{3c1}
A & = & A_\mu(\rho,x) dx^\mu
\no \\
ds^2 & = & {d\rho^2 \over 4\rho^2} +g_{\mu\nu}(\rho,x) dx^\mu dx^\nu
\eea
and their expansion in large $\rho$ takes the following form,
\bea
\label{3c2}
A_\mu(\rho,x)& = & A^{(0)}_\mu(x)+{1\over \rho} A^{(2)}_{\mu}(x) +\cdots
\no \\
g_{\mu\nu}(\rho,x) & = & \rho g^{(0)}_{\mu\nu}(x) + g^{(2)}_{\mu\nu}(x)
+ {1 \over \rho} g^{(4)}_{\mu\nu}(x)
+{\ln \rho \over \rho}  g^{({\rm ln})} _{\mu\nu}(x)  + \cdots
\eea
The coefficients $g^{(4)}_{\mu\nu}$, $g^{({\rm ln})} _{\mu\nu}$, and the trace of $g^{(4)} _{\mu \nu}$
are fixed by the Einstein equations to be local functionals of the conformal boundary metric $g^{(0)}_{\mu\nu}$.
The boundary stress tensor $T^{\mu \nu}$ and current $J^\mu$ of \cite{Balasubramanian:1999re,deHaroSkenderis}
are defined in terms of the variation of the on-shell action
with respect to $g^{(0)}_{\mu \nu}$ and $A^{(0)} _\mu$ respectively (see \cite{D'Hoker:2010hr}).
In terms of the Fefferman-Graham data the result is,
\bea
\label{3c3}
4\pi G_5 T_{\mu\nu}(x) & = & g^{(4)}_{\mu\nu}(x) +{\rm local}
\no \\
2\pi G_5 J_\mu(x) & = & A^{(2)}_\mu(x)+{\rm local}
\eea
Indices are raised and lowered using the conformal boundary metric
$g^{(0)}_{\mu\nu}$.  The local terms denote tensors constructed locally
from $g^{(0)} _{\mu\nu}$ and $A^{(0)}_\mu$, which may be dropped
when computing correlators at non-coincident points.

%%%%%%%%%%%%%%%%%%%%%%%%%%%%%%%%%%%%%%%%%%%%%%%%%%%%
%%%%%%%%%%%%%%%%%%%%%%%%%%%%%%%%%%%%%%%%%%%%%%%%%%%%
\section{The purely magnetic brane: zero charge density}
\label{sec:4}
%%%%%%%%%%%%%%%%%%%%%%%%%%%%%%%%%%%%%%%%%%%%%%%%%%%%
%%%%%%%%%%%%%%%%%%%%%%%%%%%%%%%%%%%%%%%%%%%%%%%%%%%%

The case of vanishing charge density, with zero or non-zero temperature, provides a
physically interesting system, which lends itself to much simpler treatment than
the charged case. For this reason, we shall investigate it first here, in its own right.

\subsection{The purely magnetic brane at $T=0$}
\label{sec:4a}

We begin with the case of zero temperature. For vanishing charge density and temperature,
Lorentz invariance in the $t,x_3$ directions is restored. To exhibit this symmetry,
it will be convenient to re-interpret $t,x_3$ as light-cone coordinates by substituting $t \to x^+$
and $x_3 \to x^-$. In this space-time coordinate system, Lorentz transformations act by
$x^\pm \to \lambda ^{\pm 1} x^\pm$, and further restrict the Ansatz to $E=P=M=N=0$.
In terms of the remaining functions $K,L$, the Einstein equations reduce to,
\bea
\label{4a1}
0 & = & (L^2 K)'' - 24 K
\no \\
0 & = & K^2 L'' + K K' L' + 2 KK'' L - L (K')^2
\no \\
- 4 B^2 & = &  (K')^2 L^2 + 4 KK' LL' + K^2 (L')^2 - 24 K^2
\eea
Here, the prime stands for the derivative with respect to $r$.
The last equation of (\ref{4a1})  is a constraint, whose derivative with respect to $r$
is linear in the first two equations, and vanishes, as soon as these equations are satisfied.
The first equation  gives $L$ as a function of $K$ by quadrature only.
Substituting this form of $L$  into the constraint
then gives an equation for $K$ which may be solved numerically.

\sm

The reduced field equations (\ref{4a1}) admit an exact solution for which $K$ is independent of $r$,
which is given by,
\bea
\label{4a3}
K(r) = {B \over \sqrt{3}}
\hskip 1in
L(r) = 2 \sqrt{3} \, r
\eea
Substitution into the metric of (\ref{3b3}) reveals that the corresponding space-time  has
the form $AdS_3 \times {\bf R}^2$, with an $AdS_3 $ radius given by $\ell _3 =1/\sqrt{3}$.
The $AdS_5$ vacuum, with $K, L \sim r$ is not a solution to the reduced equations (\ref{4a1})
when $B\not=0$, but it does become an approximate solution in the limit of large $r$, namely
when $B/r \to 0$. The $T=0$ purely magnetic brane is the solution to (\ref{4a1}) for which the functions
$K$ and $L$ tend to the $AdS_3 \times {\bf R}^2$ solution of (\ref{4a3}) when $r \to 0$, and
is asymptotic to $AdS_5$ of (\ref{2b1}) when $r \to \infty$,
\bea
\label{4a4}
K(r) \sim c_V r
\hskip 1in L(r) \sim 2r
\eea
Numerical analysis confirms that such a purely magnetic brane solution exists and is regular
for all $r>0$, and gives the numerical value $c_V = 2.797$.

\begin{figure}[htb]
\sidecaption[t]
\includegraphics[scale=.52]{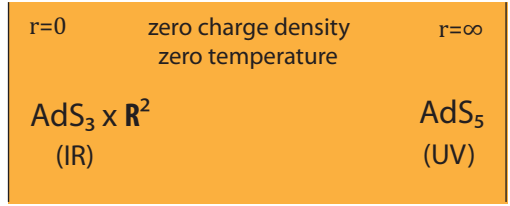}
\caption{The purely magnetic brane solution  interpolates between an $AdS_5$ space-time with magnetic
field for large $r$, and a $AdS_3 \times R^2$ space-time with magnetic field for small~$r$. The holographic
dual field theory has zero charge density and temperature.}
\label{fig:3}
\end{figure}

\subsection{RG flow and thermodynamics}
\label{sec:4b}

The holographic dual to the purely magnetic brane solution is a renormalization group flow
from 3+1-dimensional ${\cal N}=4$ super-Yang-Mills theory in the UV (for large $r$) to
a 1+1-dim. CFT in the IR (for small $r$). This flow is schematically represented in
figure \ref{fig:3}. The holographic picture  is consistent with the qualitative gauge dynamics
behavior of strongly interacting Weyl fermions discussed in section \ref{sec:2}. The central
charge $c$ of this CFT may be derived using the Brown-Henneaux formula \cite{Brown:1986nw},
\bea
\label{4b1}
c = { 3 \ell _3 \over 2 G_3} \hskip 1in {1 \over G_3} = { B \, V_2 \over G_5}
\eea
applied to an $AdS_3$ of radius $\ell_3=1/\sqrt{3}$, and where we have taken $x_{1,2}$ to be
compactified on a ${\bf T^2}$ with area $V_2$.

\sm

The specific heat coefficient at low temperature may be expressed in terms of the entropy
density $s$ by $s/T$.
In turn, the Cardy formula gives the entropy density $s$ in terms of the central charge of
a 1+1-dimensional CFT.
It may be used here to extract the holographic specific heat coefficient $s_{\rm grav} /T$, and the
entropy density $s_{\rm grav}$ in terms of the zero temperature purely magnetic brane, and we find,
\bea
\label{4b2}
{s_{\rm grav} \over T} = { \pi \over 3} \, c  = { \pi B \, V_2 \over 2 \sqrt{3} G_5}
= \sqrt{{4 \over 3}} \, {s_{\rm gauge} \over T}
\eea
In the last equality, we have included the comparison with the entropy density $s_{\rm gauge}$
evaluated earlier for free fermions in the lowest Landau level. To exhibit this relation, we have
used the AdS/CFT relation $G_5=\pi/(2N^2)$. The fact that the gravity and gauge theory
central charges do not agree can be understood as follows \cite{Almuhairi:2011ws}.
Comparing the central charges at small (but finite) and large values of the 't Hooft
coupling should show agreement, because the central charge of a D=1+1 CFT is
unchanged under marginal deformation.  However, the passage from zero to small 't Hooft
coupling can be a discontinuous change if the CFT has a relevant operator that is either
absent or present in the two cases.   The $\sqrt{4/3}$ factor is presumably a result of the
appearance of this  relevant operator.

\sm

To derive the thermodynamics of the purely magnetic brane at all $T$, we replace
the near-horizon $AdS_3 \times {\bf R}^2$ space by a BTZ $\times {\bf R}^2$ black brane.
The latter is expected to solve the field equations (\ref{3b1}) as well since BTZ may be
obtained as a quotient of $AdS_3$ by a discrete group. Concretely, the absence of electric
charge allow us to set $E=P=0$ in (\ref{3b3}), but the fields $M,N$ need to be retained at finite
temperature. The gauge field is still $F = B dx^1 \wedge dx^2$, while the metric takes the following form,
\bea
\label{4b3}
K(r)= {B \over \sqrt{3}}
& \hskip 1in &
L(r) = 4 \sqrt{3} (r-r_+)
\no \\
N(r)=1~~~
&&
M(r) = - 12 (r-r_+)(r_+-r_-)
\eea
The Hawking temperature is found to be $T=3(r_+-r_-)/\pi$. Numerical analysis confirms the existence
of a regular solution that interpolates between the above near-horizon BTZ $\times {\bf R}^2$
solution for small $r$, and asymptotically $AdS_5$ for large $r$. Using this pure magnetic brane
solution for arbitrary $T$, the entropy density may be calculated at all $T$, and is found to behave as
$s_{\rm grav} \sim T^3$ for high $T$, with the standard factor of 3/4 compared to the high $T$ gauge
theory calculation.

\subsection{Calculation of current-current correlators at $T=0$}
\label{sec:4c}

The boundary current formalism discussed in section \ref{sec:3c} may be used to evaluate
the various two-point functions of the gauge current ${\cal J}^\mu$ and the stress tensor
${\cal T}^{\mu \nu}$. In the absence of charge density for the purely magnetic brane,
the cross correlators $\< {\cal J}^\mu (x) {\cal T}^{\mu \nu}(y)\>$ will vanish identically.
We begin by evaluating the current-current correlators, by combining linear response theory
with the formulas of (\ref{3c3}),
\bea
\label{4c1}
J^\mu (x) = i \int d^4 y \, \sqrt{- \det (g^{(0)}_{\mu \nu}) }
\Big \< {\cal J} ^\mu (x) {\cal J}^\nu (y) \Big \> \delta A^{(0)} _\nu (y)
\eea
Here, the expectation value $J^\mu$ of the current ${\cal J}^\mu$ is sourced by a linear variation
in the source $A^{(0)}_\nu$ of the gauge potential, using (\ref{3c3}). As no variation of
the metric is imposed, the current $J^\mu$ may be obtained by linearizing the
Maxwell-Chern-Simons equations on the first line of (\ref{3b1}) around the purely magnetic brane.
We shall be interested in correlators in the 1+1-dimensional effective low energy CFT only,
and thus restrict to excitations carrying momentum along the magnetic field direction.
The gauge potential for definite momentum is then given by,
\bea
\label{4c2}
A = A_B + \left ( a_+ (r,p) dx^+ + a_- (r,p) dx^- \right ) \, e^{ipx}
\eea
where $dA_B=B dx_1 \wedge dx_2$, and we shall use the notations $px= p_+ x^+ + p_- x^-$
and $p^2 = p_+ p_-$ throughout.
Denoting the metric fields of the purely magnetic brane solution at $T=0$ by $K$ and $L$,
we find that the Maxwell-Chern-Simons equations may be decoupled in terms of the
variables $\ep _\pm = p_- a_+ \pm p_+ a_-$ for which we obtain the following equations,
\bea
\label{4c3}
0 & = & KL  (KL \ep_-')' - 4k^2 B^2  \ep _- - 2 K^2 {p^2\over L}  \ep_-
\no \\
0 & = & KL \ep _+ ' - 2 kB \ep_-
\eea
Given that the functions $K$ and $L$ are known only numerically, solving the above
equations for general $p^2$ can only be achieved numerically. If we restrict attention
to the regime of small $p^2$, however, then the linearized equations can be solved
essentially analytically using the method of overlapping expansions.

\subsection{Method of overlapping expansions}
\label{sec:4d}

The characteristic scale of the purely magnetic brane solution, namely where the functional
dependence of $K$ and $L$ transits from the behavior of (\ref{4a4}) at large $r$
to the behavior of (\ref{4a3}) at small $r$,  is set by $r \sim 1$.

\sm

In the {\sl near-region}, defined by $r \ll 1$, the purely magnetic brane solution may be
approximated by the behavior in (\ref{4a3}), so that (\ref{4c3}) becomes,
\bea
\label{4d1}
0 & = & r^2 \ep _- '' + r \ep _- ' - k^2 \ep _- -{p^2 \over 12 \sqrt{3} r}  \, \ep _-
\no \\
0 & = & r \ep _+' - k \ep _-
\eea
The first equation is of the modified Bessel type and is solved by the Bessel functions $I_{2k}(p/\sqrt{r})$
and $K_{2k}(p/\sqrt{r})$. Only the solution $\ep _- (r) \sim K_{2k}(p/\sqrt{r})$ is regular as $r \to 0$,
which leads us to reject the solution $I_{2k}$.

\sm

In the {\sl far-region}, defined by $p^2 \ll r$, we may neglect the last term of the first equation
in (\ref{4c3}), and solve the remaining equation in terms of the function,
\bea
\label{4d2}
\psi (r) \equiv \int _\infty ^r { dr' \over K(r')L(r')}
\eea
Note that $\psi(r)$ depends only on the data of the purely magnetic brane solution.
Expressing the solution directly in terms of the original variables $a_\pm (r,p)$, we find,
\bea
\label{4d3}
a_\pm = p_\pm \tilde a_0 + \left ( a_\pm ^{(0)} - p_\pm \tilde a_0 \right ) e^{\pm 2kB \psi (r)}
\eea
where $a_\pm ^{(0)}$ and $\tilde a_0$ are integration constants.

\sm

An {\sl overlap-region}, in which the near-region and the far-region overlap in a
finite interval, will exist provided $p^2 \ll 1$. Assuming that $p^2\ll 1$, there will exist an
overlap region in which we may match the $p^2/r \ll 1$ behavior of the Bessel function
in the near-region solution,
\bea
\label{4d4}
a_\pm  = C \, {  (p^2/12)^{\mp k}  \over \Gamma (1 \mp 2k) p_\mp} \, r^{\pm k}  + p_\pm a_0
\eea
with the $r\ll 1$ behavior of the far-region solution. The latter may be derived from the asymptotic
behavior of the function $\psi (r)$, which is found to be for $r \ll 1$,
\bea
\label{4d5}
\psi (r) \sim { 1 \over 2 B} \ln r + \psi _0
\eea
Numerical evaluation gives  $\psi _0 \approx 0.2625$.
Comparing the $r$-dependence in (\ref{4d4}) and (\ref{4d3}) using (\ref{4d5}), we see that
the near-region and far-region functional behaviors are indeed the same, and given by a constant
term, as well as by $r^{\pm k}$ terms. Matching these functional dependences produces the full solution.

\subsection{Current two-point correlators}
\label{sec:4e}

To complete the calculation of the current-current correlators in the
long-distance approximation  $p^2\ll1$ we use the overlapping expansion results
obtained earlier. The large $r$ approximation for the function $\psi (r)$ is
obtained analytically from the asymptotic behavior of $K$ and $L$ in (\ref{4a4}),
\bea
\label{4e1}
\psi (r) \sim -{ 1 \over 2 c_V r}
\eea
and used to derive the asymptotic behavior for the gauge potential for $r \to \infty$,
\bea
\label{4e2}
a_\pm (r,p) = a^{(0)} _\pm +{ 1 \over 4r} a_\pm ^{(2)}
\hskip 0.8in
a^{(2)} _\pm = \mp { 4 kB \over c_V} \left ( a^{(0)} _\pm - p _\pm \tilde a_0 \right )
\eea
To obtain $\tilde a_0$ and $a^{(2)}_\pm$ in terms of $a_0$ and $a^{(0)}_\pm$,
we match the near-region solution of (\ref{4d4}) with the far-region solution of (\ref{4d3}).
Including proper normalizations \cite{D'Hoker:2010hr}, we obtain the
current-current correlators in the limit $p^2 \ll 1$,
\bea
\label{4e3}
\Big \< {\cal J} _+  (p) {\cal J}_+ (-p) \Big \> & = & { kc \over 2 \pi} \, { p_+ \over p_-} \, { 1 \over 1- \zeta p^{4k}}
\no \\
\Big \< {\cal J} _-  (p) {\cal J}_- (-p) \Big \> & = & { kc \over 2 \pi} \, { p_- \over p_+} \, { \zeta p^{4k} \over 1- \zeta p^{4k}}
\no \\
\Big \< {\cal J} _+  (p) {\cal J}_- (-p) \Big \> & = & - { kc \over 2 \pi}  \, { \zeta p^{4k} \over 1- \zeta p^{4k}}
\eea
where $c$ is the Brown-Henneaux central charge derived in (\ref{4b1}),
and $\zeta = \zeta (k)$ is a $k$-dependent function whose precise form will not be needed here.
We note that the above correlators saturate the chiral anomaly relation independently of $\zeta$,
\bea
\label{4e4}
p_+ {\cal J}_- + p_- {\cal J}_+ = {kc \over \pi} \left ( p_+ a_- ^{(0)} - p_- a_+ ^{(0)} \right )
\eea
To leading order in small $p^2$ the correlators involving ${\cal J}_-$ both vanish, while the
correlator involving only ${\cal J}_+$ takes the following form in position space,
\bea
\label{4e5}
\Big \< {\cal J} _+  (x) {\cal J}_+ (0 ) \Big \> = -{kc \over 2 \pi^2} \, {1 \over (x^+)^2}
\eea
With our conventions, the above sign in the central term corresponds to a unitary Abelian Kac-Moody
algebra for $k>0$ and $c >0$, as is the case here.

\subsection{Maxwell-Chern-Simons holography in $AdS_3$}
\label{sec:4f}

Attempts to formulate Maxwell-Chern-Simons (MCS) holography directly in $AdS_3$ space-time
are fraught with subtleties \cite{Andrade:2011sx}.
This circumstance may be investigated directly with the help
of the near-region solutions derived in section \ref{sec:4d}, and in particular the
large $r$ asymptotics of this solution given in (\ref{4d4}),
\bea
\label{4f1}
a_\pm (r) = \alpha _\pm + \beta_\pm  r^{\pm k}
\eea
From the point of view directly of $AdS_3$, it is unclear which coefficients should be used as
sources, and which ones correspond to expectation values.   That is, it is not clear which
boundary conditions lead to a consistent theory, and indeed most boundary conditions lead to
problems  with instabilities and/or ghosts  \cite{Andrade:2011sx}.   Symptoms of this may be
detected in the
current-current correlators derived in (\ref{4e4}), by taking  the limit $p^2 \gg 1$. In this limit,
the correlators involving the component ${\cal J} _+$ vanish, while the correlator of ${\cal J}_-$
becomes in position space,
\bea
\label{4f2}
\Big \< {\cal J} _-  (x) {\cal J}_- (0 ) \Big \> = +{kc \over 2 \pi^2} \, {1 \over (x^-)^2}
\eea
Although this correlator  by itself saturates the chiral anomaly, its sign corresponds to that
of a {\sl non-unitary} Abelian Kac-Moody algebra. This violation of unitarity is a symptom of the
disease which besets certain choices of boundary conditions for  Maxwell-Chern-Simons directly in $AdS_3$.

\sm

However, by the same token we see that when the MCS theory is obtained
as the IR limit of the holographic RG flow provided by the purely magnetic brane solution
from an asymptotic $AdS_5$ completion, then the MCS theory makes perfect sense.
The key point is that the IR theory comes with a built in UV cutoff, given by the scale at
which the AdS$_3$ factor goes over to AdS$_5$.  All the would-be inconsistencies are
removed by the presence of the UV cutoff.

\subsection{Effective conformal field theory and double-trace operators}
\label{sec:4g}

The leading IR contribution to the two-point function of ${\cal J}_+$ in (\ref{4e3})
may be parametrized by an effective free scalar field $\phi$ with canonical Lagrangian
${\cal L} _\phi \sim \p_+ \phi  \p_- \phi$. The subdominant $p^{4k}$ terms in the
correlators of (\ref{4e3}) may be understood in terms of contributions to the Lagrangian
from double-trace operators ${\cal L} _{\cal O} \sim \p_+ {\cal O} \p_- {\cal O}$,
where ${\cal O}$ is a conformal primary field of dimension $(k,k)$.   The expressions for the currents
and two-point functions (in momentum space) of these operators are given as follows,
\bea
{\cal J}_+ = \p_+ \phi + \p_+ {\cal O}
& \hskip 0.6in &
\< \phi (p) \phi (-p) \> = p^{-2}
\no \\
{\cal J}_- =  \p_- {\cal O} \hskip 0.39in
& \hskip 0.6in &
\< {\cal O} (p) {\cal O}(-p) \> = \zeta p^{4k-2}
\eea
These two fields summarize the entire IR behavior of the correlators, within our approximations.
The importance of double trace operators in the holographic renormalization group
\cite{Balasubramanian:1999jd,de Boer:1999xf} has been stressed recently in
\cite{Heemskerk:2010hk,Faulkner:2010jy}.

\sm

We note that the structure of the ${\cal J}_+$ correlators is quite reminiscent
of the structure of the charge density two-point function in the Luttinger liquid model.
Clearly there is a leading inverse square contribution both in (\ref{4e5}) and in
(\ref{2b3}), while the the $p^{4k}$ higher order corrections in (\ref{4e3}) are
analogous to the $x^{-2\Delta}$ corrections of (\ref{2b3}).

\subsection{Stress tensor correlators and emergent Virasoro symmetry}
\label{sec:4h}

Since the purely magnetic brane produces a flow from $AdS_5$ towards a space-time
containing an $AdS_3$ factor, its holographic dual is expected to be a full-fledged CFT
in the IR limit, endowed with left- and right-moving Virasoro algebras. This structure, and
the value of the associated Brown-Henneaux central charge $c$ of (\ref{4b1}) dictate the
structure of the two-point function of two stress tensor components.  All correlators involving
the component ${\cal T}_{+-}$ vanish at non-coincident points, as does the mixed correlator
$\< {\cal T} _{++} (x) {\cal T}_{--}(y)\>$. The remaining correlators are given by,
\bea
\label{4h1}
\left \< {\cal T}_{\pm \pm } (x) {\cal T}_{\pm \pm } (0) \right \> = { c \over 8 \pi ^2 (x^\pm)^4}
\eea
These two-point correlators may be checked by explicit calculation using the method of
overlapping expansions along the same lines as for the current correlators, and agree.

\sm

The existence of two Virasoro symmetry algebras in the IR brings to light the holographic
realization of the  {\sl emergence of symmetries}. Indeed, the Brown-Henneaux
coordinate transformations on the near-horizon $AdS_3$, which produce these Virasoro
asymptotic symmetry algebras, correspond to pure gauge transformations.  This is as expected,
since gravity in three space-time dimensions is (locally) trivial.
But these coordinate transformations on $AdS_3$ extend to perturbative deformations of the
pure magnetic brane solution and interpolate to the $AdS_5$ boundary where they
correspond to physical deformations which are not merely gauge transformations.
Indeed, the asymptotic symmetry algebra at the asymptotically $AdS_5$ boundary of the purely
magnetic brane is $SO(4,2)$, a finite-dimensional Lie algebra of which the infinite-dimensional
Virasoros are certainly not subalgebras. Therefore, we conclude that the
Virasoro symmetries present in the IR are {\sl emergent symmetries}, not present in the UV theory.

%%%%%%%%%%%%%%%%%%%%%%%%%%%%%%%%%%%%%%%%%%%%%%%%%%%%
%%%%%%%%%%%%%%%%%%%%%%%%%%%%%%%%%%%%%%%%%%%%%%%%%%%%
\section{Holographic Dual Solutions for non-Zero Charge Density}
\label{sec:5}
%%%%%%%%%%%%%%%%%%%%%%%%%%%%%%%%%%%%%%%%%%%%%%%%%%%%
%%%%%%%%%%%%%%%%%%%%%%%%%%%%%%%%%%%%%%%%%%%%%%%%%%%%

A non-zero charge density gives rise to a wealth of interesting physics. As discussed in the
introduction, the physical location of the charge, namely either inside or outside the event
horizon and a mixture thereof, will to a large extent govern the phase diagram of the dual
field theory. Remarkably, it will be possible to understand most of the low temperature
dynamics, for large enough magnetic field, using the analytical methods of overlapping expansions,
supplemented by a few numerical constants determined from the purely magnetic brane solution.
In this section, we shall proceed analytically, and fill in the regions of the phase diagram not
accessible through analytical results with the help of numerical results.

\subsection{Reduced field equations}
\label{sec:5a}

Investigating thermodynamics in the presence of charge density and a magnetic field
in the $x^3$ direction will involve gravitational solutions which are invariant under translations
in $x^\mu$, and rotations in the $x_1, x_2$ plane. Thus, we need the full
Ansatz of equation (\ref{3b3}). The corresponding reduced field equations are as follows,
\bea
\label{5a1}
{\rm M1} &&\quad  \left((NE+LP)e^{2V}\right)'+2kbP =0
\\ \no
{\rm M2} && \quad    \left((LE+MP)e^{2V}\right)'-2kbE =0
\\ \no
{\rm E1} && \quad L''+2V'L' +4(V'' +V'^2)L -4PE=0
\\ \no
{\rm E2} && \quad M'' +2V'M'+4(V'' +V'^2)M +4E^2=0
\\ \no
{\rm E3} && \quad N''+2V'N'+4(V'' +V'^2)N+4P^2  =0
\\ \no
{\rm E4}  && \quad  f (V')^2 +  f' V'  +{1 \over 4}(L')^2 - {1 \over 4} M'N' + b^2 e^{-4V}
+ MP^2 + 2 LEP + NE^2 = 6
\\ \no
{\rm fV} && \quad (f e^{2V})'' = 24\, e^{2V}
\eea
We have used the notation $f=L^2-MN$ of (\ref{3b4}), and changed variables to $K=e^{2V}$.  Also, we now denote the magnetic field as $b$, reserving the use of $B$ for the value of the magnetic field in a canonical coordinate system.

\sm

The reduced field equations admit a number of first integrals. Using the potentials
$A$ and $C$ for $E=A'$ and $P=-C'$, equations M1 and M2 admit obvious first integrals,
\bea
\label{5a2}
(NA'-LC')e^{2V} + 2 kb C & = & 0
\no \\
(LA'-MC')e^{2V} - 2 kb A & = & 0
\eea
The integration constants to $A$ and $C$ that arise here have been absorbed into the
definition of these functions. Forming combinations of equations E1, E2, and E3, and using
(\ref{5a2}), we find the following further first integrals,
\bea
\label{5a3}
\lambda e^{2V} - 4 kb AC = \lambda _0 & \hskip 1in & 2 \lambda = NM' - MN'
\no \\
\mu e^{2V} + 4 kb A^2 = \mu _0 & \hskip 1in & ~ \mu = LM' - ML'
\no \\
\nu e^{2V} + 4 kb C^2 = \nu _0 & \hskip 1in & ~ \nu = NL' - LN'
\eea
where  $\lambda _0, \mu _0$, and $\nu_0$ are the constant values of the
corresponding first integrals. Equation fV is linear in $f$ and may be solved for as a
function of $V$. Finally, $\lambda, \mu, \nu$ satisfy a  purely kinematic relation,
\bea
\label{radial}
(f')^2 =  4 (\lambda ^2 - \mu \nu) + 4f  (L')^2 - 4 f M'N'
\eea
Under constant $ \Lambda \in SL(2,{\bf R})$ transformations of the coordinates~$x^\pm$,
\bea
\label{5a2a}
\left ( \matrix{ \tilde x^+ \cr \tilde x^- } \right ) = \Lambda ^{-1} \left ( \matrix{ x^+ \cr x^- } \right )
\eea
the Ansatz (\ref{3b3}), the reduced field equations (\ref{5a1}), and the first integrals (\ref{5a2})
and (\ref{5a3})  are invariant provided the fields transform as,
\bea
\label{5a3a}
\left ( \matrix{ \tilde A \cr - \tilde C } \right ) = \Lambda ^t
\left ( \matrix{ A \cr -C } \right )
\hskip 0.7in
\left ( \matrix{ \tilde M & \tilde L \cr \tilde L & \tilde N } \right ) = \Lambda ^t
\left ( \matrix{ M & L \cr L & N } \right ) \Lambda
\eea
while the field $V$ and the combination $f$ are invariant. The triplet $(\lambda, \mu, \nu)$ is the
$SL(2,{\bf R})$ analogue of angular momentum and transforms under the vector
representation of $SL(2,{\bf R})$, just as their first integral values  $(\lambda _0, \mu_0, \nu_0)$ do.

\subsection{Near-horizon Schr\"odinger geometry}
\label{sec:5b}

Introducing charge requires that $E \not=0$ in the Ansatz of (\ref{3b3}). Re-interpreting $t$ as $x^+$
and $x_3$ as $x^-$, we see that turning on a charge corresponds to a deformation which is null in the
$x^\pm$ coordinate system. This suggests the existence of a solution in which deformations
in the $x^-$ directions vanish. We begin by exhibiting an exact charged near-horizon solution
in which the $x_1,x_2$-directions are frozen out by the presence of a magnetic field, so that
$K=e^{2V}$ is constant. The gauge potential and electric field are found as follows,
\bea
\label{5c1}
A(r) ={e_0 \over k} \, r^k
\hskip 0.7in
E(r) = e_0 \, r ^{k-1}
\eea
The near-horizon metric takes the form,
\bea
\label{5c2}
ds^2 = { dr^2 \over 12 r^2} + 4 \sqrt{3} r dt dx_3
- \left ( \alpha _0 r +{ 2 e_0^2 \, r^{2k} \over k(2k-1)} \right ) dt^2 + dx_1^2 + dx_2^2
\eea
In these coordinates we have $b=\sqrt{3}$.
This metric coincides with the Schr\"odinger space-time of   \cite{Balasubramanian:2008dm,Son:2008ye}
and  the null-warped solution of \cite{Anninos:2010pm}.

\subsection{The charged magnetic brane solution}
\label{sec:5c}

The near-horizon Schr\"odinger geometry at $r \to 0$ extends to a regular {\sl charged magnetic brane
solution} to the full reduced field equations (\ref{5a1}) with asymptotic $AdS_5$ behavior.
In this respect, the role played by the Schr\"odinger near-horizon geometry for the charged
magnetic brane is parallel to the role played by the $AdS_3 \times {\bf R}^2$ near-horizon
geometry of the purely magnetic brane. The full solution has vanishing deformations
in the $x^-$ direction, so that we can set $C=N=0$, and the functions $K=e^{2V}$ and $L$
remain those of the purely magnetic brane. The remaining fields $A$ and $M$ may be obtained
by quadrature from $K=e^{2V}$ and $L$, and we find,
\bea
\label{5b1}
A(r) & = & A_\infty \, e^{2kb \psi (r)}
\no \\
M(r) & = & L(r) \left ( - {\alpha _\infty \over 2 \sqrt{3}}
- 4kb \int ^r _\infty  { dr' \, A(r')^2 \over K(r') L(r')^2} \right )
\eea
The function $\psi (r)$ was defined in (\ref{4d2}). Using the $r \to 0$ asymptotics
of $\psi (r)$ given in (\ref{4d5}), we see that the gauge potential satisfies the standard regularity
condition $A(0)=0$ at the horizon. Using the $r \to \infty$ asymptotics of (\ref{4e1}), we see that
the integration constant $A_\infty= A(\infty)$ is the chemical potential. The integration constant
$\alpha _\infty$ introduces a relative tilt between the light-cones in the UV and the IR.
Solutions for different values of $A_\infty$ and $\alpha_\infty$ are related to one another by
$SL(2,{\bf R}) $ transformations which preserve the restrictions $C=N=0$, and we have,
\bea
\label{5b3}
\Lambda = \left ( \matrix{\lambda _1 & 0 \cr \lambda _2 & \lambda _1 ^{-1} \cr } \right )
\hskip 1in
\matrix{ \tilde A_\infty  & = & \lambda _1 A_\infty   \cr
\tilde \alpha_\infty  & = & \lambda _1 ^2 \alpha _\infty - 2 \lambda _1 \lambda _2  \cr }
\eea
Therefore, all solutions with $A_\infty \not= 0$ are equivalent to one another under $SL(2,{\bf R}) $.

\sm

The asymptotic behavior near the boundary of $AdS_5$ is given as follows,
\bea
\label{5b2}
A(r) & \sim &  A_\infty - { c_E \over r} \hskip 1in c_E = { kb \, A_\infty  \over  c_V}
\no \\
M(r) & \sim & - { \alpha_\infty \over \sqrt{3}}   r
\eea
The integration constants $e_0$ and $ \alpha _0$ of the near-horizon Schr\"odinger geometry
may be related to the parameters $\alpha _\infty$ and $A_\infty$ of the boundary, and we find,
\bea
\label{5c3}
A_\infty &  = & e_0 \, e^{-2kb \psi _0}
\no \\
\alpha _\infty & = & \alpha _0 + 16 c_V^2 c_E^2 J(k)
\hskip 0.7in J(k) ={1 \over 2k} \int _0 ^\infty dr { e^{4kb \psi (r)} \over K(r) L(r)^2}
\eea
where the constant $\psi _0$ was defined in (\ref{4d5}).

\subsection{Regularity of the solutions}
\label{sec:5d}

In anticipation of extending the charged magnetic brane solution to finite
temperature, we must require regularity of the solution as a  black brane.
Thus, the coefficient function of $dt^2$ in the metric must remain negative
throughout the space-time region outside the (outer) horizon, which leads
us to require,
\bea
\label{5d1}
M(r) \leq 0
\eea
with equality only at the horizon $r=0$. At infinity, this imposes the condition $\alpha _\infty >0$.
The solution near the horizon of (\ref{5c2}) imposes further conditions which depend on the
value of $k$. In the parameter region $0 \leq k < 1/2$, the $r^{2k}$ term dominates over
the $\alpha _0 r$ term, and leads to $M(r) >0$ as soon as $e_0 \not= 0$. Thus, the
charged magnetic brane solution  in the region $0 \leq k < 1/2$ is excluded, as it cannot arise as the zero temperature limit of a nonsingular finite temperature black brane.

\sm

In the parameter region $1/2 < k$, it is the $\alpha _0 r$ term that dominates,
which requires $\alpha _0 \geq 0$. The value $\alpha _0 =0$ is actually regular
as well, since the $r^{2k}$ term contributes negatively for $1/2 <k$. It is straightforward
to see from (\ref{5b1}) that these conditions are also sufficient to make the
charged magnetic brane solution regular for all $0<r<\infty$.

\subsection{Existence of a critical magnetic field}
\label{sec:5e}

The regularity conditions derived in the preceding section on the parameters $\alpha _0$
and $\alpha _\infty$ may be translated into conditions on physically observable parameters
in the dual field theory. Since the boundary field theory is conformal invariant, only
dimensionless combinations of data can enjoy physical meaning. The magnetic field
$B$ and the charge density $\rho$ have non-trivial dimension, but the ratio defined by,
\bea
\label{5e1}
\hat B \equiv { B \over \rho ^{2/3}}
\eea
is dimensionless, and physically observable. Expressions for $B$ and $\rho$, which denote the values of the magnetic field and charge density in coordinates such that the  AdS$_5$ metric takes a canonical for,
may themselves may be read off from the boundary behavior of the solution, and are
given by,
\bea
\label{5e2}
B = { 2 b \over  c_V}
\hskip 1in
\rho = 4 c_E \sqrt{{2 b \over \alpha _\infty}}
\eea
where $c_E$ was defined in (\ref{5b2}). Thus, $\hat B^3$ may be
cast in the following form,
\bea
\label{5e3}
\hat B^3 = { 3 \alpha _\infty \over 4 c_V^3 c_E^2}
\hskip 1in
\hat B _c^3 \equiv { 3 (\alpha _\infty - \alpha _0) \over 4 c_V^3 c_E^2} = { 12 J(k) \over c_V}
\eea
Here, we have also defined the combination $\hat B_c$ in terms of which
we obtain the following final expression for $\hat B$,
\bea
\label{5e4}
{\hat B^3_c \over \hat B^3} = 1 - { \alpha _0 \over \alpha _\infty}
\eea
Positivity of $J(k)$ for $k >0$ implies $ \alpha _0 < \alpha _\infty$, while
regularity required $0 \leq \alpha _0$. Thus, we conclude that the charged
magnetic brane solution obtained above is regular if and only if $\hat B_c \leq \hat B$.
In this sense, $\hat B _c$ represents a critical magnetic field. Its value depends
only on the CS coupling $k$ and  the data of the purely magnetic brane solution.
Inspection of the behavior of $L$ and $\psi$ in the integral for $J(k)$ shows
that $J(k)$, and hence $\hat B_c$ diverges as  $k \to 1/2$, thus providing a natural
physical end point for the validity of the charged magnetic brane solution.

\subsection{Low $T$ thermodynamics for $\hat B > \hat B_c$}
\label{sec:5f}

The low $T$ behavior dual to the charged magnetic black brane solution must be
investigated separately for magnetic fields $\hat B > \hat B_c$ and $\hat B \sim \hat B_c$.
We begin here with the study of the former. The presence of a low non-zero temperature
induces only small changes to the charged magnetic brane solution for large $r$, while
substantially altering its near-horizon behavior. The corresponding leading $T$-dependent
near-horizon behavior needs to be treated exactly to incorporate these effects.

\sm

Our starting point is the purely magnetic BTZ$\times {\bf R}^2$ solution
already discussed in section \ref{sec:4b}. Its metric is given by (\ref{4b3}), but it
will be convenient here to choose the outer horizon at $r=0$, so that $r_+=0$,
and to parametrize the solution as follows,
\bea
\label{5f1}
F & = & b dx^1 \wedge dx^2
\no \\
ds^2 & = & { dr^2 \over 12 r^2 +mnr} - mr dt^2 + 4 \sqrt{3} dtdx_3 + ndx_3^2 + dx_1^2 + dx_2^2
\eea
For an asymptotically $AdS_5$ space-time given by,
\bea
\label{5f3}
ds^2 = {dr^2 \over 4r^2} - {\alpha _\infty \over \sqrt{3}} r dt^2 + 4r dt dx_3 + c_V (dx_1^2 + dx_2^2)
\eea
the dimensionless form of the entropy density $\hat s$, and of the temperature $\hat T$ may
be expressed as follows,
\bea
\label{5f2}
\hat s \equiv { s \over B^{3/2}} = {\sqrt{n c_V \alpha _\infty} \over 24}
\hskip 0.8in
\hat T \equiv { T \over B^{1/2}} = { m \sqrt{n c_V} \over 4 \pi \sqrt{\alpha _\infty}}
\eea
In their ratio all reference to $n$ cancels out,
\bea
\label{5f4}
{ \hat s \over \hat T} = {\pi \over 6} \, {\alpha _\infty \over m}
\eea
This ratio has a finite limit as  $T \to 0$, and may be evaluated in terms of the data of the
$T=0$ charged magnetic solution, for which we have $m = \alpha _0$ and $n=0$ by (\ref{5c2}).
Along with the result for $\alpha _0/\alpha _\infty$ from (\ref{5e4}), we find
a remarkably simple formula,
\bea
\label{5f5}
{ \hat s \over \hat T} = {\pi \over 6} \, {\hat B^3 \over \hat B^3 - \hat B_c^3}
\eea
A number of remarks are in order.
\begin{enumerate}
\itemsep=0in
\item In our system, the physical entropy density vanishes at zero temperature (in contrast with
the non-vanishing entropy density used in \cite{Liu:2009dm,Faulkner:2009wj,Cubrovic:2009ye} in 2+1 dimensions.
\item The limit $\hat B \to \infty$ corresponds to vanishing charge density $\rho$ at fixed $B$,
and reproduces the zero charge density result of (\ref{4b2}).
\item The dependence on $\hat B / \hat B_c$ manifested in (\ref{5f5}) is reminiscent of the
dependence on the excitation velocity and couplings in the Luttinger liquid theory in  (\ref{2b3}).
\item The divergence of $\hat s / \hat T$ at zero temperature as $\hat B \to \hat B_c$ signals
the presence of a quantum critical point at $\hat B_c$. Therefore, in a small region around
$T=0$ and $\hat B=\hat B_c$ in the $\hat T,\hat B$ plane, we should expect to find quantum
critical behavior, to be explored in the subsequent subsections.
\item The phase for $\hat B < \hat B_c$ has non-zero entropy density at $\hat T=0$,
and may be thought of as a deformation of the Reissner-Nordstrom solution for zero
magnetic field.
\item Numerical solutions perfectly reproduce the above analytical approximations,
as will be explained in section  \ref{sec:5j}.

\end{enumerate}

\subsection{Low $T$ thermodynamics for $\hat B = \hat B_c$}
\label{sec:5g}

Precisely at the quantum critical point, we have $\hat B= \hat B_c$, or equivalently $\alpha _0=0$.
The resulting near-horizon metric of (\ref{5c2}) then simplifies slightly,
\bea
\label{5g1}
ds^2 = { dr^2 \over 12 r^2} + 4 \sqrt{3}r dt dx_3
- { 2 e_0^2 \, r^{2k} \over k(2k-1)} dt^2 + dx_1^2 + dx_2^2
\eea
More importantly, however, the metric now is invariant under the
following scaling transformations,
\bea
\label{5g2}
r \to \lambda r \hskip 0.8in t \to \lambda ^{-k} t \hskip 0.8in x_3 \to \lambda ^{k-1} x_3
\eea
with $x_1, x_2$ unchanged. The associated dynamical scaling exponent is given by,
\bea
\label{5g3}
z= { k \over 1-k}
\eea
General arguments show that, for fixed $\hat B = \hat B_c$, the entropy
density scales with temperature according to  the relation $\hat s \sim \hat T^{d/z}= T^{1/z}$
given that the scaling theory here has space dimension $d=1$.

\sm

Numerical analysis shows that the above prediction, based on the structure of the
near-horizon metric and its scaling symmetry, is borne out in only a limited range for $k$.
The actual behavior is as follows,
\bea
\label{5g4}
\hat s \sim  \hat T ^{(1-k)/k}
& \hskip 1in &
1/2  < k < 3/4
\no \\
\hat s \sim  \hat T ^{1/3} \hskip 0.2in
& \hskip 1in &
3/4  < k
\eea
The numerical accuracy for the exponents is better than 1 \% for the points we have checked.
Note that the exponent matches continuously across the value $k=3/4$.
Inspection of the numerically obtained metric functions shows that the near-horizon
region for finite $\hat T$ holds an electrically charged black brane whose space-time
metric differs from that of BTZ.

\sm

To understand why the arguments based on the scaling symmetry of the near-horizon metric
fail for $k>3/4$, we consider a scaling transformation which leaves the $AdS_3$ part
of the metric invariant, but not necessarily the term in $e_0^2$. For example,
applying the following scaling,
\bea
\label{5g5}
r \to \lambda r \hskip 0.8in t \to  t/\sqrt{\lambda} \hskip 0.8in x_3 \to x_3 / \sqrt{\lambda}
\eea
with $x_1, x_2$ unchanged, will scale the term in $e_0^2$ by a factor of $\lambda ^{2k-1}$.
Scaling towards the IR corresponds to $\lambda <1$, and we see that the term in $e_0^2$
naively becomes irrelevant. Whether the term indeed is irrelevant becomes a dynamical
question, which is not easy to settle. Detailed arguments were given in \cite{D'Hoker:2010ij} that
the separation point is indeed $k=3/4$. The scaling exponent of 1/3 may be reproduced
for the range $k>1$ with the help of the method of overlapping expansions.
The calculations are technically involved, and will not be reproduced here.

\sm

\begin{figure}[htb]
\sidecaption[t]
\includegraphics[scale=.52]{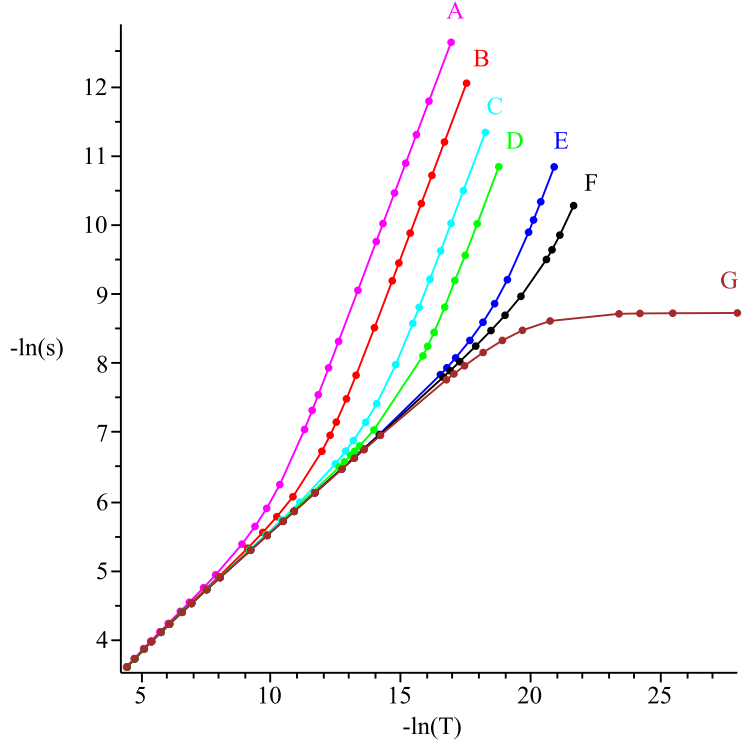}
\caption{The cross-over behavior of $\hat s$ versus $\hat T$ for $k=k_{\rm \scriptstyle susy} = 2/\sqrt{3}$
at values of $\hat B > \hat B_c$ corresponding to the curves
$A: ~ \hat B^3 = 0.125$,
$B: ~ \hat B^3 = 0.1247$,
$C: ~ \hat B^3 = 0.1246$,
$D: ~ \hat B^3 = 0.12458$,
$E: ~ \hat B^3 = 0.12457$,
$F: ~ \hat B^3 = 0.124569$, and
$G: ~ \hat B^3 = 0.124568$.
To lighten notations, hats on the variables
$ \hat T, \hat s$ have not been exhibited in labeling the figure.
At moderately low temperatures, $\hat s$ scales as $\hat T^{1/3}$ (lower left corner),
while at ultra-low temperatures $\hat s$ scales as $\hat T$ for $\hat B > \hat B_c$
(curves $A,B,C,D,E,F$) and tends to a non-zero constant for $\hat B < \hat B_c$
(curve $G$). The dots represent numerical data points, while the solid interpolating
lines are included to guide the eye. }
\label{fig:5}
\end{figure}

In figure \ref{fig:5} we display numerical data illustrating the crossover from the
behavior $\hat{s} \sim \hat{T}$ to the behavior $\hat{s} \sim \hat{T}^{1/3}$ for
$k=k_{\rm \scriptstyle susy} =2/\sqrt{3}$. Qualitatively, the cross-over behavior
of figure \ref{fig:5} persists for all $k > 3/4$.

\subsection{Scaling function in the quantum critical region}
\label{sec:5h}

In a small region surrounding the quantum critical point $T=0$ and $\hat B = \hat B_c$,
a critical scaling regime sets in. For the range $k >1$, we have been able to derive
this scaling behavior with the help of the method of overlapping expansions, and we find,
\bea
\hat s = \hat T^{1/3} \, f \left ( { \hat B - \hat B_c \over \hat T^{2/3}} \right )
\label{scalef}
\eea
for a certain scaling function $f$ (which is not to be confused with the metric function $f$
introduced in (\ref{5a1})). At $\hat B = \hat B_c$, this formula reproduces the
scaling behavior discussed in (\ref{5g4}) of the previous section for $k>1$.
For $\hat B > \hat B_c$, and low temperature, namely $\hat T^{2/3} \ll (\hat B - \hat B_c)$,
we should recover (\ref{5f5}), so that we should have $f(x) \sim \pi \hat B_c / (18 x)$
for large $x$. Actually, the method of overlapping expansions allows one to compute
$f(x)$ for the range $k >1$, and we shall quote here the result without reproducing the derivation
given in \cite{D'Hoker:2010ij},
\bea
f(x) \left ( f(x)^2 + { x \over 32 k \hat B_c^4} \right ) = { \pi \over 576 k \hat B_c ^3}
\eea
The scaling function $f$ continues to apply for $\hat B < \hat B_c$,
and we find the following behavior for the entropy density as a function of the
magnetic field,
\bea
\hat s = {\sqrt {\hat B_c - \hat B} \over 4 \sqrt{2k} \, \hat B_c ^2}
\eea
The value of the exponent is reproduced to approximately 0.2 \% accuracy
by numerical simulations, and the prefactor to approximately 20 \%.

\subsection{Numerical completion of the holographic phase diagram}
\label{sec:5j}

\begin{figure}[htb]
\sidecaption[t]
\includegraphics[scale=.6]{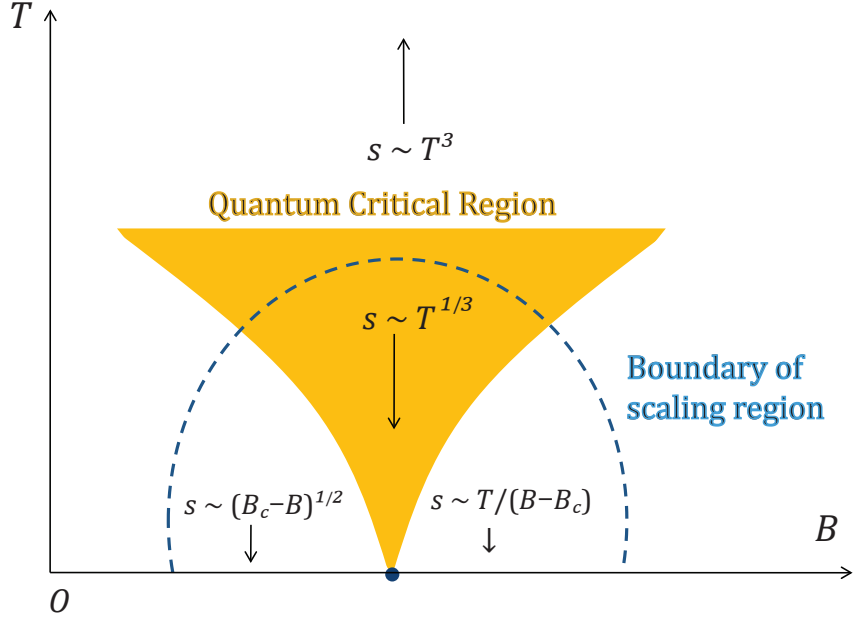}
\caption{The full holographic phase diagram in terms of the variables $\hat B$ and $\hat T$
for $k>3/4$.  Hats on $\hat B, \hat T, \hat s$, and $\hat B_c$
have not been exhibited in the figure.}
\label{fig:4}
\end{figure}

The full holographic phase diagram in the variables $\hat B, \hat T$ is presented in figure \ref{fig:4}.
Here, the various asymptotic behaviors are combined onto a single graph for the range
$k > 3/4$. For the range $1/2 < k < 3/4$, the scaling exponent 1/3 at $\hat B = \hat B_c$
must be replaced by $(1-k)/k$, and the behavior of the entropy density for $\hat B < \hat B_c$
is altered though we have not systematically studied the corresponding modifications.

\subsection{Correlators at non-zero charge density}
\label{sec:5k}

Correlators of the Maxwell current ${\cal J}^\mu$ and of the stress tensor ${\cal T}^{\mu \nu}$
may be evaluated, in the long distance approximation, in the presence of a magnetic field
at zero temperature, but now with
non-vanishing background electric charge density $\rho$, or equivalently, with chemical
potential $\mu$. As in the case with vanishing charge density studied in sections \ref{sec:4c},
 \ref{sec:4e}, and \ref{sec:4h}, we use the method of overlapping expansions of \ref{sec:4d}
 valid for $k >1$.
The calculations for the charged case proceed in analogy with the ones for the neutral
case, but are now considerably more  delicate and technically involved. We refer to
the original paper \cite{D'Hoker:2010ij} for their detailed derivation, and restrict here to
quoting and explaining the results.

\sm

Correlators involving the operators with minus chirality are {\sl unmodified} from the zero
charge case.  In particular, the two point function of ${\cal J}_-$ has no singularities,
while the two point function of ${\cal T}_{--}$ continues to be given by (\ref{4h1}).
The correlators with plus chirality are found as follows,
\bea
\label{5k1}
\Big \< {\cal J} _+  (x) {\cal J}_+ (0 ) \Big \> & = & -{kc \over 2 \pi^2} \, {1 \over (x^+)^2}
\no \\
\Big \< {\cal J} _+  (x) {\cal T}_{++} (0 ) \Big \> & = & +{kc \mu \over 2 \pi^2} \, {1 \over (x^+)^2}
\no \\
\Big \< {\cal T} _{++}  (x) {\cal T}_{++} (0 ) \Big \> & = & - {kc \mu^2 \over 2 \pi^2} \, {1 \over (x^+)^2}
+ { c \over 8 \pi ^2 (x^+)^4}
\eea
where $\mu $ is the chemical potential, related to the charge density by,
\bea
\label{5k2}
\mu = A_\infty = {\rho c_V \sqrt{\alpha _\infty} \over 4 kb \sqrt{2b}}
\eea
The system of correlators in the presence of charge may be related to the system of
correlators of operators ${\cal J}_+ ^{(0)} $ and ${\cal T}_{++} ^{(0)} $ at zero charge
density by the following simple operator mixing relations,
\bea
{\cal J }_+ & = & {\cal J}_+ ^{(0)}
\no \\
{\cal T}_{++} & = & {\cal T}_{++} ^{(0)} - \mu _+ {\cal J}_+ ^{(0)}
\eea
Here, we have exhibited the natural Lorentz weight of the chemical potential by setting $\mu=\mu_+$.
We see that the underlying Abelian Kac-Moody algebra for ${\cal J}_+^{(0)} $
and the underlying Virasoro algebras for ${\cal T}_{\pm \pm} ^{(0)}$ are unmodified,
with unchanged Kac-Moody level $kc$, and Virasoro central charge $c$.

\subsection{Comments on stability}

The solutions studied here can, at least for special values of $k$, be uplifted to full solutions of higher dimensional supergravity and string theory, but nothing guarantees that they are stable solutions. There are two types of potential instabilities to be aware of: those coming from fields already included in our analysis, and those required by a consistent embedding into supergravity/string theory.  Regarding the former, it has been observed in several contexts that the combination of electric charge and Chern-Simons terms can lead to instabilities towards spatially modulated phases  \cite{Domokos:2007kt,Nakamura:2009tf}.  In some cases new solutions with reduced symmetry can be found \cite{Donos:2011qt,Iizuka:2012iv,Donos:2012gg,Donos:2012gg,Iizuka:2012wt}.
As for the latter, a supergravity/string theory embedding will typically bring along a variety of charged fields, and these may be unstable towards forming a condensate, as in holographic superconductors \cite{Hartnoll:2008vx}.    It is clearly an important challenge to determine when our solutions are unstable, and if not, to characterize the nature of the true ground state.

%%%%%%%%%%%%%%%%%%%%%%%%%%%%%%%%%%%%%%%%%%%%%%%%%%%%
%%%%%%%%%%%%%%%%%%%%%%%%%%%%%%%%%%%%%%%%%%%%%%%%%%%%
\section{Quantum Criticality in 2+1 dimensions}
\label{sec:7}
%%%%%%%%%%%%%%%%%%%%%%%%%%%%%%%%%%%%%%%%%%%%%%%%%%%%
%%%%%%%%%%%%%%%%%%%%%%%%%%%%%%%%%%%%%%%%%%%%%%%%%%%%

The critical theories studied so far originate from an underlying 3+1-dimensional
gauge theory in the UV which flows towards an effective 1+1-dimensional strongly interacting
CFT in the IR. Low temperature thermodynamics and long-distance correlators all
signal massless propagation along the direction parallel to the magnetic field only.
In the gravity dual, this IR behavior results from the existence of a near-horizon
Schr\"odinger geometry of the form WAdS$_3 \times {\bf R}^2$, where WAdS$_3$ is a {\sl null-warped}
deformation of $AdS_3$ space-time. The physical mechanism driving the quantum critical
transition on the gravity side is the gradual expulsion of electric charge from the
inside of the black brane horizon to the outside of the horizon as the magnetic field
$\hat B=B/\rho ^{2/3}$ is being increased; see \cite{Hartnoll:2011pp}\ for another example
of this phenomenon.

\sm

While quantum criticality in 1+1 dimensions is certainly of considerable physical interest,
as was pointed out in the preceding section, it is probably even more urgent to extend
the study to higher dimensions. Quantum criticality in 2+1 dimensions is relevant to the
physics of layered materials, such as cuprates, and graphene. In the present section,
we shall exhibit quantum critical behavior in 2+1 dimensions systems in the presence
of a magnetic field, and a non-vanishing electric charge density by holographic methods.
Criticality here is driven by the same holographic mechanism that governed the 1+1-dimensional
case, namely charge expulsion from the black brane horizon. This time, however,
the IR behavior in the gravity dual results from a flow from $AdS_6$ in the UV
to a near-horizon Lifshitz geometry \cite{Kachru:2008yh} in the IR which is a deformation
of $AdS_4$.  See \cite{Goldstein:2009cv,Braviner:2011kz} for other examples of holographic
RG flows involving Lifshitz spacetime.  The AdS$_6$ geometry in the UV should be thought
of as being dual to some 5+1 dimensional CFT, examples of which do exist,
and have been identified in \cite{Seiberg:1996bd}.

\subsection{Field equations and structure of the solutions}

The charge expulsion mechanism operating in the flow from $AdS_5$ to deformations
of $AdS_3$ is made possible by the presence of the Chern-Simons interaction for the
Maxwell field, and the existence of the transition crucially depends upon the strength
of the associated Chern-Simons coupling $k$. This is because the Chern-Simons term
provides the mechanism by which the bulk gauge field can carry its own charge.

\sm

The charge expulsion mechanism in higher dimensions that we shall focus on will
also be made possible by the presence of Chern-Simons terms. Starting with $AdS_6$
in the UV does not support a Chern-Simons terms for the bulk gauge field all by itself.
Thus, we are led to introducing further form fields. In the simplest extension,
we add a single two-form potential $C$ with field strength $G=dC$.
The corresponding Einstein-Maxwell-Chern-Simons action then becomes,
\bea
\label{7a1}
S & = & - { 1 \over 16 \pi G_6} \int d^6x \, \sqrt{g}
\left ( R - {20 \over \ell^2}  + F_{MN} F^{MN} +{ 1 \over 3} G_{MNP} G^{MNP}  \right )
+ S_{CS}
\no \\
S_{CS} & = &  { k \over 4 \pi G_6} \int C \wedge F \wedge F
\eea
where $G_6$ is the 5+1-dimensional Newton constant, $F=dA$ is the Maxwell field strength,
and $-20/\ell^2 $ stands for the cosmological constant for an asymptotic $AdS_6$ vacuum solution
of radius $\ell$, which we shall set to 1. Boundary and counter term contributions to the
action are not being exhibited here.

\sm

The Maxwell-Chern-Simons field equations are,
\bea
\label{7a2}
d * F - 2 k F \wedge G & = & 0
\no \\
d * G + k F \wedge F & = & 0
\eea
while the Einstein equations are,
\bea
\label{7a3}
R_{MN} = - 2 F_{MP} F_N{}^P - G_{MPQ} G_N {}^{PQ} +g_{MN} \left ( 5 + {1 \over 4}
F_{PQ} F^{PQ} + {1 \over 6} G_{PQR} G^{PQR} \right )
\eea
Clearly, the charge densities for both the $F$ and $G$ fields
are proportional to the Chern-Simons coupling $k$.

\sm

As we focus here on thermodynamic questions, we shall be interested in solutions
which are invariant under translations in $x^\mu = (t, x_1, x_2, x_3,x_4)$. The flow from
$AdS_6$ in the UV to $AdS_4$ and its deformations in the IR will be generated by
a constant magnetic field, which we shall choose in the direction $F_{34}=B$.
It is natural to require rotation invariance in the $x_3,x_4$ plane, as well as in the
remaining space directions $x_1,x_2$. A general Ansatz invariant under these symmetries
was constructed in \cite{D'Hoker:2012ej}, and is given by,
\bea
\label{7a4}
F & = & E dr \wedge dt + \tilde B dx_1 \wedge dx_2 + B dx_3 \wedge dx_4
\no \\
G & = & (G_1 dr + G_2 dt) \wedge dx_1 \wedge dx_2
+ (G_3 dr + G_4 dt) \wedge dx_3 \wedge dx_4
\no \\
ds^2 & = & {dr^2 \over U} - U dt^2 + e^{2V_1} (dx_1^2 + dx_2^2) + e^{2V_2} (dx_3^2 + dx_4^2)
\eea
By translation invariance in $x^\mu$, the coefficients $B, \tilde B, E, G_1, G_2, G_3, G_4, U, V_1, V_2$ depend only on $r$.
By the Bianchi identities for $F$ and $G$, the quantities $B, \tilde B, G_2, G_4$ must actually
be independent of $r$. The magnetic field $\tilde B$ plays the role of a magnetic field
living in the IR 2+1-dimensional field theory, and will be set to zero here for simplicity, $\tilde B=0$.
For $k B \not=0$, the field equations then imply that $G_2=0$ and $G_3G_4=0$. Solutions with
either $G_3 \not =0$ or $G_4 \not=0$ do not have regular horizons, so we set also $G_3=G_4=0$.
The remaining reduced field equations were derived in \cite{D'Hoker:2012ej}, and will not be
repeated here as they are reasonably involved.

\subsection{Horizon and asymptotic data, physical quantities}

We choose a coordinate $r$ such that the horizon is at $r=0$. We normalize the scales
of the coordinates $x^\mu$ by setting,
\bea
U(0)=V_1(0)=V_2(0) =0 \hskip 1in U'(0)=1
\eea
The field equations relate the horizon values $G_1(0)= -2kb E(0)$.
The asymptotic behavior for $r \to \infty$
may be parametrized analogously,
\bea
U(r)  \sim r^2
\hskip 0.8in
e ^{2V_1(r)} \sim v_1 r^2
\hskip 0.8in
e ^{2V_2(r)} \sim v_2r^2
\eea
The asymptotics of the gauge field strength fixes the physical charge density $\rho$ of the
boundary theory by $r^4 E(r) \to \rho $. The dimensionless magnetic field $\hat B = B/\sqrt{\rho}$,
temperature $\hat T=T/\sqrt{B}$, and entropy density $\hat s = s/B^2$ are then given by,
\bea
\hat B = { b \over v_2 \sqrt{\rho}}
\hskip 0.8in
\hat T = { \sqrt{v_2} \over 4 \pi \sqrt{b}}
\hskip 0.8in
\hat s = { v_2 \over 4 v_1 b^2}
\eea
The equation of state corresponds to the relation $\hat s = \hat s (k, \hat T , \hat B)$.
We shall begin by discussing below analytical solutions available in various limits.
Obtaining the function $\hat s$ throughout parameter space will, however, require numerical analysis.

\subsection{Flows towards the electric IR fixed point}

In the absence of a magnetic field, $B=0$, the purely electric solution is given by the
standard Reissner-Nordstrom form,
\bea
U= r^2 + {q^2 \over 6 r^6} -{M\over r^3}
\hskip 0.7in
V_1=V_2= \ln r
\hskip 0.7in
E={\rho \over r^4}
\eea
In the extremal limit, the location of the horizon $r_+$ is determined by $U(r_+)=U'(r_+)=0$,
and the entropy density $s \sim \rho \sim r_+^4$ does not vanish at $T=0$.
The near-horizon geometry of the purely electric solution is $AdS_2 \times {\bf R}^4$.

\sm

For $B \not =0$, numerical analysis confirms the existence of a charged magnetic brane
solution whose near-horizon  behavior coincides with that of the purely electric solution,
provided the Chern-Simons coupling $k$ remains below a critical value $ k_c$ which will be
determined shortly.

\subsection{Flows towards the  magnetic IR fixed point}

The near-horizon behavior of the purely magnetic solution is given by
$AdS_4 \times {\bf R}^2$ space-time at $T=0$, or an $AdS_4$ Schwarzschild
solution at $T\not=0$.  The two cases may be described  together by,
\bea
U={20 \over 9} \left ( r^2 -{r_+^3 \over r} \right )
\hskip 0.6in
e^{2V_1}= {20 \over 9} r^2
\hskip 0.6in
e^{2V_2} = \sqrt{{3 \over 10}} B
\eea
The temperature behaves as $T \sim r_+$, and the entropy density may be computed
exactly in the low $T$ approximation,
\bea
s = {\pi ^2 \over 5} \, \sqrt{3 \over 10} \, B T^2
\eea
This $T$-dependence is precisely as expected of a 2+1-dimensional CFT
associated with the near-horizon $AdS_4$ space-time.

\sm

For $\rho \not= 0$, numerical analysis confirms
the existence of a charged magnetic brane solution whose near-horizon
behavior is that of the purely magnetic solution, provided $k$ is
larger than the critical value $k_c$ already identified to end the purely electric flow.
The $T^2$-dependence of the entropy density which is
characteristic of 2+1-dimensional CFT behavior,  persists as long as $k > k_c$,
but the coefficient is now found to be a non-trivial function,
\bea
s = A(k,\hat B) \, B \, T^2
\eea
We expect that the problem of calculating the function $A(k,\hat B)$ may be amenable to
analytical treatment, especially since numerical evaluation indicates the
following behavior for intermediate and small $\hat B$,
\bea
A(k, \hat B) \sim c(k) \, \exp \left ( d(k) \hat B^{-2} \right )
\eea
with the characteristic behavior $d(k) \sim (k^2-k_c^2)^{-1}$ obeyed to remarkable accuracy.

\subsection{Flows towards the Lifshitz IR fixed point}

The near-horizon geometries, $AdS_2 \times {\bf R}^4$  for $k < k_c$, and
$AdS_4 \times {\bf R}^2$ for $k > k_c$, are separated
by a Lifshitz near-horizon geometry at the critical point $k=k_c$. Seeking
near-horizon solutions which are invariant under space-time scalings,
$r \to \lambda r$, ~ $t \to t/ \lambda$, ~ $x_{1,2} \to \lambda ^{-1/z} x_{1,2}$,~
and  $x_{3,4} \to \lambda ^{-\beta} x_{3,4}$, for real constants $z$ and $ \beta$,
we find that the existence of such a solution requires $\beta =0$,
as well as
\bea
k = k_c = {1 \over \sqrt{3}}
\eea
The dynamical scaling exponent $z$ is constrained to the range $z>1$,
but otherwise arbitrary. From this scaling behavior in the near-horizon region, the
scaling behavior of the low temperature behavior of the entropy density may be deduced,
and we find,
\bea
\hat s \sim T^{2 \over z}
\eea
in accord with the space-dimension of the IR theory being $d=2$, and the
parameter $z$ standing for the dynamical critical exponent.  As $z$ runs through
the range $1 < z < \infty$, the entropy density is being interpolated from its IR
behavior for the purely electric fixed point at $z=\infty$ to its IR behavior for the
purely magnetic fixed point at $z=1$. This interpolating behavior is reflected
in the functional dependence of $z$ on the magnetic field $\hat B$, which
interpolates between the following asymptotic behaviors,
\bea
{1 \over z} = \left \{ \matrix{
0.105 \, \hat B^2 & \hat B \ll 1 \cr
1- 0.894 \, \hat B^{-4} & \hat B \gg 1 \cr} \right .
\eea

\subsection{The full phase diagram}

The full phase diagram may now be assembled from the behavior in the
various regimes that we have examined in the preceding sections. For
large $T$, the entropy density is dominated by temperature alone, and
charge density $\rho$ as well as magnetic field $B$ have negligible effects.
Thus, we have $s \sim T^4$, as expected form the dual field theory by scaling.

\sm

For $k<k_c$, the flow from $AdS_6$ in the UV is towards the Reissner-Nordstrom type
near-horizon geometry $AdS_2 \times {\bf R}^4$ in the IR with $s \not=0$ at $T=0$.
Without doubt, this holographic solution will become unstable once charged scalar fields,
and/or space-dependent modulations are allowed. For $k>k_c$, the flow from $AdS_6$ in
the UV is towards the $AdS_4 \times {\bf R}^2$ near-horizon geometry of the purely
magnetic brane, with its entropy density behaving as $s \sim B T^2$, characteristic
of scaling in a 2+1-dimensional CFT. Finally, in the critical region, where $k \sim k_c$,
the entropy density at low $\hat T$ is governed by by a scaling function,
\bea
\hat s = f(k-k_c , \hat B, \hat T)
\eea
In the absence to date of a (semi-) analytical solution connecting the near-horizon
Lifshitz geometry to the asymptotic $AdS_6$ region, the scaling function
$f(k-k_c, \hat B, \hat T)$ is accessible only through numerical analysis.

\begin{figure}[htb]
\sidecaption[t]
\includegraphics[scale=.5]{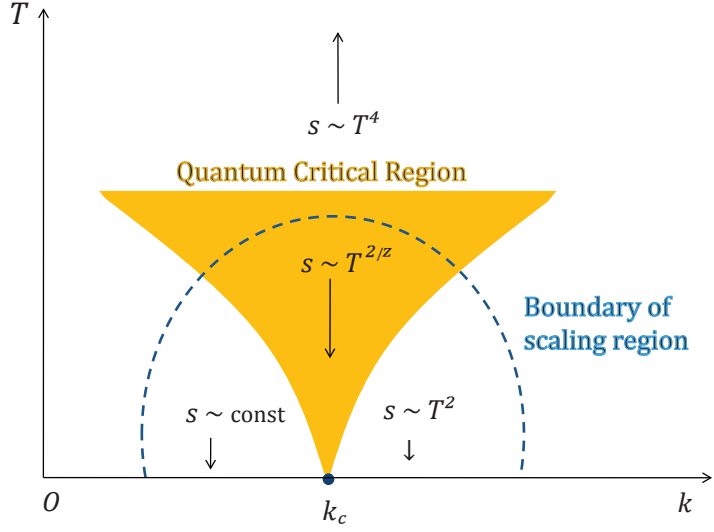}
\caption{The full holographic phase diagram in terms of the variables $k$ and $\hat T$.
To lighten notations, hats on $ \hat T, \hat s$ have not been exhibited in the figure.}
\label{fig:8}
\end{figure}

%%%%%%%%%%%%%%%%%%%%%%%%%%%%%%%%%%%%%%%%%%%%%%%%%%%%
%%%%%%%%%%%%%%%%%%%%%%%%%%%%%%%%%%%%%%%%%%%%%%%%%%%%
\section{Relation with quantum criticality in condensed matter}
\label{sec:6}
%%%%%%%%%%%%%%%%%%%%%%%%%%%%%%%%%%%%%%%%%%%%%%%%%%%%
%%%%%%%%%%%%%%%%%%%%%%%%%%%%%%%%%%%%%%%%%%%%%%%%%%%%

In this section, we shall point towards some exciting, though still speculative,
applications of our holographic results to problems in condensed matter physics.
One application of the 1+1-dimensional quantum criticality problem studied
above is to Strontium Ruthenates.

\subsection{Meta-magnetic transitions in Strontium Ruthenates}

The phase transitions exhibited by our holographic systems occur at a finite value of the magnetic field and involve no change of symmetry.   These are referred to as {\sl meta-magnetic phase transitions}.  A typical experimental situation is the following.    A material has a line of first order phase transitions at finite temperature, reached by dialing the magnetic field, and with the line ending at a finite temperature critical point.  By tuning some other control parameter, one can attempt to bring this critical point down to zero temperature, resulting in a quantum critical meta-magnetic transition \cite{Millis02}, analogous to what we have found holographically.  In particular, thermodynamic quantities such as the specific heat will diverge as the critical magnetic field is approached.

A version of this behavior, with some interesting twists, occurs in the Strontium Ruthenate compound Sr$_3$Ru$_2$O$_7$, and has been the subject of much experimental and theoretical interest in the past few
years; e.g.  \cite{RPMMG,FKLEM}. Sr$_3$Ru$_2$O$_7$ is a layered material, which,
for a large magnetic field perpendicular to the 2-dimensional layers (around 8 T) exhibits meta-magnetic behavior with a ``shrouded" quantum critical point.  Notably, as the magnetic field reaches its critical value  the entropy density behaves as $ s/  T \sim 1/(B-B_c)$, just as  we found in our AdS$_5$ system.  There appears to be no satisfactory theoretical understanding of this behavior.   While this divergence in the entropy density appears to signal the onset of a quantum critical point, what actually seems to happen  \cite{OKE,FKLEM} is that the system evolves into a nematic phase for $7.8 T < B < 8.1 T$.  Spatial anistropy in the nematic phase can be detected by applying a small in-plane component of magnetic field, which acts to align the domains, and then looking for anistropic behavior of transport coefficients.    The nematic phase seems to shroud the quantum critical point in a manner analogous to what occurs in high temperature superconductors.    The fact that the would be divergence in $s/T$ is cutoff by the appearance of a nematic phase has been described as nature's solution to the problem of avoiding a non-zero entropy density at zero temperature.

\begin{figure}[htb]
\sidecaption[t]
\includegraphics[scale=.4]{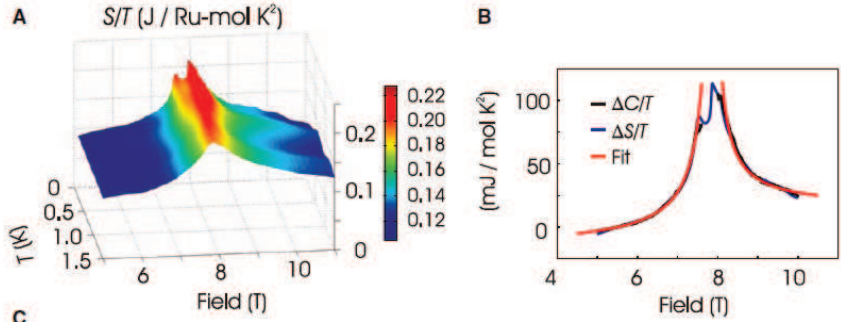}
\caption{Entropy landscape of Sr$_3$Ru$_2$O$_7$ near the meta-magnetic transition.   The right panel illustrates the $1/(B-B_c)$ divergence of $S/T$, which is ultimately cut off by the appearance of a nematic phase.   Figure taken from \cite{RPMMG}.}
\label{fig:6}
\end{figure}

In \cite{RPMMG}, the complete {\sl entropy landscape}
of Sr$_3$Ru$_2$O$_7$ at finite temperature and magnetic field has been mapped out.    The parallels with our system are clear, namely the $1/(B-B_c)$ divergence in the entropy density to temperature ratio; see figure \ref{fig:6}.  The most obvious difference is that our system is effectively 1+1 dimensional at the critical point, while Sr$_3$Ru$_2$O$_7$ is strongly 2+1 dimensional.    But it is interesting to speculate whether a nematic phase will occur also in our holographic setup.   In our ansatz we have assumed full translational and rotational invariance, but recent results indicate that there are frequently  instabilities towards anisotropic phases \cite{Domokos:2007kt,Nakamura:2009tf,Donos:2011qt,Iizuka:2012iv,Donos:2012gg,Donos:2012gg,Iizuka:2012wt}.

\subsection{Relation to Hertz-Millis theory}

A standard approach to modeling magnetically tuned quantum phase transitions is based on the Hertz-Millis theory \cite{hertz,millis93,LRMW}.  In d+1 dimensions we consider the effective action
\bea S = \int\! d\omega d^d k \left( {|\omega| \over |k|} +k^2 +(\hat{B}- \hat{B}_c)\right) |\phi(\omega,k)|^2 + \ldots~.
\label{HMaction}
\eea
The bosonic field $\phi$ represents the local magnetization, and one is supposed to think of this action as arising from integrating out gapless fermions at one-loop.  There is no controlled approximation that justifies this approach, and indeed it is known to sometimes lead to predictions in conflict with experiment.   Let us nonetheless make the following suggestive observations. We consider the action (\ref{HMaction}) with $d=1$, and compare to our asymptotically AdS$_5$ critical theory.   At $\hat{B}=\hat{B_c}$ the Hertz-Millis
action is scale invariant, with $k$ and $\omega$ assigned scaling dimensions $1$ and $3$ respectively.  The dynamical critical exponent is therefore $z=3$, which matches our AdS$_5$ result for $k> 3/4$, and will lead to the scaling law for the entropy, $s\sim T^{1/3}$, in one spatial dimension, as we found.  Furthermore, $\hat{B} - \hat{B}_c$ plays the role of a relevant coupling of scaling dimension 2.    This agrees with (\ref{scalef}); to see this note that the argument of $f$  in (\ref{scalef}) has vanishing scale dimension, and $\hat{T}$ shares the same scaling dimension as $\omega$, namely $3$.    Therefore, the scaling predictions of the Hertz-Millis theory can be identified in our holographic setup.  Of course there are also differences; for instance, there is no analog of our finite ground state entropy density branch for $\hat{B} \leq \hat{B}_c$.

\bigskip

\noindent
{\bf Acknowledgments}

This work was supported in part by NSF grant PHY-07-57702.

\sm

During the course of this entire project, we have benefited from helpful conversations and correspondence
with several colleagues, and we wish to thank here Vijay Balasubramanian, David Berenstein,
Sudip Chakravarty, Geoffrey Comp\`ere, Jan de Boer, Fr\'ed\'eric Denef, St\'ephane Detournay,
Tom Faulkner, Jerome Gauntlett, Sean Hartnoll, Gary Horowitz, Finn Larsen, Alex Maloney,
Eric Perlmutter, Joe Polchinski, Matt Roberts, Joan Simon, and especially Akhil Shah who
collaborated on one of our papers. During parts of this work, we have enjoyed the hospitality
of the KITP during the ``Quantum Criticality and AdS/CFT Correspondence" program in 2009, and
of the Aspen Center for Physics in 2011. One of us (E. D.) wishes to thank the Laboratoire de
Physique Th\'eorique de l'Ecole Normale Sup\'erieure, and the Laboratoire de Physqiue Th\'eorique
et Hautes Energies, CNRS and Universit\'e  Pierre et Marie Curie - Paris 6, and especially Constantin
Bachas and Jean-Bernard Zuber for their warm hospitality while part of this work was being completed.

\end{document}